# Individual utilities of life satisfaction reveal inequality aversion unrelated to political alignment

Crispin Cooper, Ana Fredrich, Tommaso Reggiani, Wouter Poortinga

## Abstract

How should well-being be prioritised in society, and what trade-offs are people willing to make between fairness and personal well-being? We investigate these questions using a stated preference experiment with a nationally representative UK sample (n = 300), in which participants evaluated life satisfaction outcomes for both themselves and others under conditions of uncertainty. Individual-level utility functions were estimated using an Expected Utility Maximisation (EUM) framework and tested for sensitivity to the overweighting of small probabilities, as characterised by Cumulative Prospect Theory (CPT).

A majority of participants displayed concave (risk-averse) utility curves and showed stronger aversion to inequality in societal life satisfaction outcomes than to personal risk. These preferences were unrelated to political alignment, suggesting a shared normative stance on fairness in well-being that cuts across ideological boundaries. The results challenge use of average life satisfaction as a policy metric, and support the development of nonlinear utility-based alternatives that more accurately reflect collective human values. Implications for public policy, well-being measurement, and the design of value-aligned AI systems are discussed.

## 1. Introduction

Life satisfaction has gained substantial traction as a metric for public policy evaluation, with much current interest in its use as a direct measure of outcomes in place of traditional economic indicators such as income or GDP (Layard, 2006; Cato, 2008; Frijters and Krekel, 2021; Carney, 2020). Advocates of the approach argue that life satisfaction captures a broader, more subjective conception of welfare that aligns more closely with what individuals value. However, life satisfaction-based policymaking faces a core conceptual challenge: life satisfaction is not fungible in the way monetary metrics are. A change from, e.g., 6 to 7 on a self-reported satisfaction scale may not reflect the same change in utility across individuals, and the scale itself lacks a natural unit that permits aggregation or comparison in cardinal terms.

Despite this, much empirical and policy-focused work treats life satisfaction scores as if they are directly comparable and additive across individuals. Researchers routinely average life satisfaction as though each unit represents a constant and interpersonally comparable increment of welfare. This practice is rarely justified and often implicit, yet it underpins a substantial body of work in well-being economics and public policy (Bronsteen, Buccafusco, and Masur, 2013; Cooper, 2020).

A further issue with averaging well-being scores is that the practice implicitly endorses a Benthamite version of utilitarianism, in which social value is equated with the total amount of ‘good’. In contrast, a Rawlsian conception of value would posit that what matters is the well-being of the least well off (Fleurbaey, 2011; Layard, 2016). These two positions lie at opposite ends of a spectrum of possible social welfare functions. In valuating policies, we must inevitably adopt a position somewhere along this spectrum, either implicitly or explicitly.

The question of how to aggregate individual-level well-being to determine social value is also highly relevant to the field of AI alignment (Everitt, Lea, and Hutter, 2018). Although primarily concerned with the question of how to ensure artificially intelligent systems are aligned with human goals, this field has also recognized the importance of better defining those goals ourselves. The possibility of capturing risk preference through nonlinear utility curves has been briefly discussed (Vamplew et al., 2018; Smith, Klassert, and Pihlakas, 2022), as has the relevance within AI of well-being and fairness (Fjeld et al., 2020; Gabriel et al., 2024; Maden, Lomas, and Hekkert, 2024). Current best practice is codified in the IEEE-7010 standard (IEEE, 2020). As such, the current study can also be considered a contribution related to Democratic AI (Koster et al., 2022).

In the context of life satisfaction, one suggested approach to selecting such a position is to consult the public on the values they prioritise (Frijters et al., 2020), a research direction also identified as a high priority by Layard and De Neve (2023).

The contribution of this paper is to address this issue empirically by asking the public how they would make trade-offs between distributions of life satisfaction under uncertainty. The aim is to estimate the utility functions individuals apply to life satisfaction when evaluating both personal risk and societal inequality in well-being, and to demonstrate the potential application of these functions. We do this through a stated preference experiment in which participants choose between different life satisfaction distributions, with varying levels of inequality and average well-being, under a “veil of uncertainty” regarding their own place within the distribution.

The specific objectives are,

(1) to validate comprehension and reliability of responses, ensuring that observed preferences reflect meaningful and interpretable trade-offs.
(2) to estimate personal and social utilities of life satisfaction, and how they vary by political alignment,
(3) to demonstrate the application of these utilities in an exploratory aggregation of UK life satisfaction data,
(4) to test sensitivity of results to application of small probability weighting.

The veil of uncertainty (or veil of ignorance) approach has a long-standing tradition in welfare economics and distributive justice (Rawls, 1971). Its central premise is that one society may be considered more just than another - even if both are unequal - if individuals would rationally choose to live in it while unaware of their eventual position within the distribution. This philosophical framework thus posits a link between individuals’ tolerance for personal risk and their acceptance of societal inequality, which we explicitly test in the current study.

By presenting participants with decisions made under uncertainty about their own position, we infer their implicit preferences regarding inequality and risk in terms of life satisfaction. We estimate individuals’ risk and inequality aversion using flexible, nonparametric models fitted to experimental standard gamble data. Modelling these preferences allows us to move beyond the simplistic assumption of additive life satisfaction and instead develop a framework that respects the ordinal and context-dependent nature of life satisfaction data.

The remainder of the paper is structured as follows: Section 2 reviews background literature including decision-making theories and attempts to compute utilities for the subjective life satisfaction scale. Section 3 explains methods for the survey and its analysis. Section 4 shows results, including validation of responses, estimation of utilities, application to aggregating UK life satisfaction data, and sensitivity test to the choice of theoretical model. Sections 5-7 cover discussion, limitations and conclusions respectively.

## 2. Literature

Of the various measures of subjective well-being, we focus on *life satisfaction*, which is the most widely accepted indicator of the construct and is most closely aligned with desired life outcomes (Fujiwara and Campbell, 2011). The measure is typically operationalised using the following question: “On a scale of 0 to 10, where 0 is “not at all” and 10 is “completely” [...] Overall, how satisfied are you with your life nowadays?” (Office for National Statistics, 2018).

A parallel development can be seen in the health domain, where Quality Adjusted Life Years (QALYs) are used to evaluate health outcomes (Brazier et al., 2017; Frijters et al., 2020). Recent work proposes the development of a comparable measure not just for health, but overall life satisfaction, termed the Well-Being Adjusted Life Year (WELLBY), which weights each year of life by a person’s life satisfaction score (Frijters et al., 2024). The WELLBY applies a linear weighting, thus implicitly taking a Benthamite position on social value.

At the core of life satisfaction valuation is the notion of utility, which can be defined as the quantity individuals seek to maximise in their decision making. Utilities can be inferred from the choices people make in experimental settings, under the assumption that participants are more likely to choose the option they perceive to have greater utility. The exact interpretation of utility, however, depends on the chosen theoretical model. Expected Utility Maximization (EUM), developed by von Neumann and Morgenstern (1953), is a normative theory describing how one should make decisions. In contrast, Cumulative Prospect Theory (CPT), developed by Kahneman and Tversky (1979; 1992), is primarily a descriptive theory of how people actually make choices, biased by loss aversion and probability weighting. CPT can also serve as a normative theory, if used to ‘de-bias’ observed choices to determine latent utilities (Bleichrodt, 2002). In a policymaking context, however, this move is delicate: asserting that people’s ‘true’ preferences differ from their stated ones is in clear conflict with the principles of direct democracy. Although representative democracies allow room for interpreting public preferences (Sumption, 2019), it would nonetheless be a bold claim to assert that one’s own statistical model of voter preferences is more accurate than what voters themselves report.

In practice, policymakers tend to use EUM rather than CPT, in particular in health (Reed Johnson et al., 2013; European Medicines Agency (EMEA), 2010; National Institute for Clinical Excellence (NICE), 2013); a state of the field also acknowledged by its critics (Abellan-Perpiñan, Bleichrodt, and Pinto-Prades, 2009; Attema, Brouwer, and l’Haridon, 2013). Importantly, both EUM and CPT theories converge on a formal definition of loss aversion: the relative weighting of a loss compared to an equal sized gain. EUM encodes this directly into the utility curve, while CPT encodes such information in valuation functions. Although CPT may correct for probability weighting, the basic comparison remains meaningful. We therefore adopt this shared definition as our measure of loss aversion and test its sensitivity to the use of a CPT-style probability weighting function.

If loss aversion is present in decision-making, then under EUM the utility function must be nonlinear. This challenges several arguments for utility linearity with respect to life satisfaction, as summarised by Frijters and Krekel (2021). First, it is argued that asking participants to rate life satisfaction on a linear scale encourages them to treat changes of ±1 as equivalent across the scale - such that a move from 8 to 9 is perceived as comparable to a move from 3 to 4. Second, it is found that prediction errors in life satisfaction are homoscedastic, implying that the likelihood of over- or underestimation is constant across the scale. Third, it is claimed that individuals can estimate the life satisfaction scores of others with reasonable accuracy.

We argue that these points conflate perceptual linearity of life satisfaction with utility linearity. Even if a life satisfaction scale is perceptually linear, it does not follow that utility is a linear function of the life satisfaction scale. If choices consistent with loss aversion are found, then under EUM the utility function is by definition nonlinear. Under CPT such choices could reflect nonlinear utilities, nonlinear utility valuation, biased probabilities, or a combination of all three.

Two notable experimental studies provide an empirical precedent to challenge the assumption of linearity. Peasgood et al. (2018) used a Time Trade-Off (TTO) method to elicit utility weights for life satisfaction, finding a mildly concave (risk-averse) utility curve. This method avoids bias from probability weighting. Mukuria et al. (2023) reported a more pronounced concave utility curve on the EQ-HWB-S health and well-being scale; although not explicitly discussed, this is visible in the fact that lower states on the perceptually linear EQ-HWB-S scale were found to have greater utility decrements (until a utility of zero was reached, beyond which, states were rated worse than dead and show the opposite). This is based on TTO, augmented by a discrete choice experiment requiring trade-off between different components of the multidimensional scale.

In contrast to the present study, Peasgood et al. (2018) did not attempt to anchor life satisfaction scale positions with interpersonally comparable vignettes, while Mukuria et al. (2023) used EQ-HWB-S rather than the life satisfaction scale. While both studies complement the current work, neither study directly presented participants with decisions on risk and inequality, which we argue should also be presented directly - as at least some proportion of any measured risk aversion may be rational, and should hence be reflected in decision making. Finally, neither study examined how such preferences relate to political ideology.

# 3. Methods

## 3.1. Sample and recruitment

The sample consisted of 300 participants, using stratified quota sampling to reflect the UK population in terms of age, sex, and political affiliation. Age and sex were included due to their known associations with risk-taking behaviour (Bonem, Ellsworth, and Gonzalez, 2015; Byrnes, Miller, and Schafer, 1999). Political affiliation was included because the study concerns distributive preferences and attitudes toward social inequality, which are likely to be shaped by ideological orientation.

Sample size requirements for the discrete choice experiment were estimated, using established heuristics. Orme's rule of thumb suggested a minimum of 125 participants, while Assele's more recent method indicated a threshold of 75 (Assele, Meulders, and Vandebroek, 2023). The final sample exceeded both recommendations to ensure representativeness. Sample characteristics are shown in Table 1.

| Sex | Female | | | | | Male | | | | | Female | Male | All |
|---|---|---|---|---|---|---|---|---|---|---|---|---|---|
| Age | 18-24 | 25-34 | 35-44 | 45-54 | 55+ | 18-24 | 25-34 | 35-44 | 45-54 | 55+ | All ages | | |
| Conservative | 1 | 4 | 4 | 4 | 16 | 1 | 3 | 3 | 4 | 14 | 29 | 25 | 54 |
| Green Party | 2 | 2 | 2 | 2 | 3 | 2 | 2 | 2 | 2 | 3 | 11 | 11 | 22 |
| Labour | 11 | 14 | 14 | 12 | 22 | 11 | 13 | 13 | 12 | 19 | 73 | 68 | 141 |
| Liberal Democrats | 1 | 3 | 3 | 2 | 5 | 1 | 2 | 2 | 2 | 5 | 14 | 12 | 26 |
| Reform UK | 1 | 2 | 2 | 3 | 12 | 1 | 1 | 1 | 3 | 11 | 20 | 17 | 37 |
| SNP | 1 | 1 | 1 | 1 | 1 | 1 | 1 | 1 | 1 | 1 | 5 | 5 | 10 |
| Other | 1 | 1 | 1 | 1 | 1 | 1 | 1 | 1 | 1 | 1 | 5 | 5 | 10 |
| All | 18 | 27 | 27 | 25 | 60 | 18 | 23 | 23 | 25 | 54 | 157 | 143 | 300 |

*Table 1 Characteristics of the sample at time of recruitment*

Participants were recruited through Prolific (www.prolific.com) and completed the survey on 9 April 2025. Prolific has been found to yield high-quality and diverse samples in recent comparative works (Douglas, Ewell, and Brauer, 2023). Prior to the main study, the survey was piloted with 20 participants under live video call supervision. This pilot served two purposes: to test the survey instrument and to provide a benchmark against which the quality of unsupervised responses in the main study could be assessed.

Standard data quality checks were included, in line with best practice in online survey research (Pyo and Maxfield, 2021; Robinson et al., 2019; Lovett et al., 2018; Casey et al., 2017; Chmielewski and Kucker, 2020). These included simple attention checks (e.g. “I am currently completing an online survey”), as well as monitoring of completion times to flag unusually fast responses that might indicate disengagement.

Political orientation was assessed in two ways: first by asking party affiliation - how participants would vote (if they were able to) in an election tomorrow; and second, through five items from the British Social Attitudes Survey capturing economic and social attitudes across the left-right spectrum (NatCen Social Research, 2017). Responses to these questions were summed to form a single political alignment variable for analysis. The survey questions are provided in the Supplemental Material.

Ethical approval was granted by the Cardiff University School of Computer Science & Informatics Research Ethics Committee, references COMSC/Ethics/2022/100b and COMSC/Ethics/2022/100e. Informed consent was recorded using online forms.

## 3.2. Vignettes

To ensure comparability in how participants interpret and use the life satisfaction scale, we employed a vignette-based approach (Angelini et al., 2014). Participants were asked to rate the life satisfaction of fictitious characters in a series of hypothetical life situations. These ratings provide insight into how each respondent applies the 0-10 life satisfaction scale, allowing us to control for individual differences in scale use.

Two distinct conditions were used:

1. **Gambles-first condition:** Participants were first asked to make gambles (see following section) between a set of life situations (i.e., the vignettes), and then subsequently rate the life satisfaction of each situation. This approach maximises separation between the decision task and the respondent's interpretation of the life satisfaction scale, thereby reducing contamination between scale use and utility elicitation.
2. **Life satisfaction-first condition:** Participants were first asked to rate the life satisfaction of each situation and then make the gambles between different life situations (i.e., the vignettes) using these ratings as reference points.

We used these different conditions to control for biases that may be unique to each approach. The life satisfaction-first condition could inadvertently encourage those with prior knowledge of probability and expected value to compute expected outcomes using an EUM framework. But by simplifying life situations to positions on a life satisfaction scale we also potentially reduce cognitive load, which is known to be crucial to maintain response efficiency in choice modelling in particular (Reed Johnson et al., 2013). Given that both approaches have distinct methodological advantages, we randomly assigned participants to one of the two conditions and later tested for systematic differences in responses across the two groups.

The vignettes (see Figure 1) served three purposes: first, to familiarise participants with the life satisfaction scale; second, to provide guidance on how the hypothetical scenarios should be interpreted, emphasising that participants may hold different values from those of the vignette character, but should nonetheless evaluate life satisfaction from the character's perspective rather than their own; and third, to establish a common reference point for gamble decisions. Each vignette described the character's career, relationships, and physical fitness. Vignettes A–C reflected generally healthy life states with varying levels of satisfaction. Since the differences among them do not stem from health conditions, all three could reasonably be assigned a weight of 1.0 within the QALY framework. In contrast, vignettes D and E described increasing struggles with mental health, with E also involving major physical health issues, implying that both would be evaluated with a lower QALY weight.

If participants provided ratings that violated logical orderings, for example, by rating a clearly better scenario lower than a worse one, they were given the opportunity to either revise their responses or to provide an explanation.

Overall, how satisfied are you with your life nowadays?

Please answer on a scale from 0 to 10, where 0 means "not at all satisfied" and 10 means "totally satisfied".

Next

---

[next screen]

In the previous question you answered 8 out of 10.

This is a position on what we call a *life satisfaction scale*. We find this scale useful as it allows each person to rate their life situation according to what is personally important to them.

We will now ask you to rate the life satisfaction you would expect for five imaginary people (A-E).

- Please assume that each person considers the important things in life to be the things shown in this table (career, relationships, physical fitness)
- Try not to assume each person has the same preferences as you do yourself
- You can assume that for all people in the table, life presents sufficient challenge to be interesting

| Person | Career | Relationships | Physical Fitness | Your guess at their life satisfaction 0-10 |
|---|---|---|---|---|
| A | I love my job, it is respected and well paid. I could not imagine having a better job. | My family and friends bring me a lot of joy, we spend a lot of time together. I am very happy with my relationship status. Life is great. | I adore sports and physical activities! I practice almost everyday and consider myself fortunate to do so much. | |
| B | I like my job. It is a good job, and reasonably paid. | I have a good connection with my family & friends, and I see them fairly often. I am happy with my relationship status. | I enjoy practicing sport in my spare time. | |
| C | I find my job a little dull, but it pays enough for essentials. | I visit my family and friends sometimes, but I wish we would have a strong connection. I am mostly happy with my relationship status. | I occasionally have the chance to play sports. I would like to do more it more often but I struggle to find the time. | |
| D | I don't enjoy my job. The pay is low, and sometimes I struggle to pay the bills. | I only occasionally visit my family and friends, and I often feel lonely. I am unhappy about my relationship status. | I rarely exercise. I would like to exercise more but I struggle to find time and motivation. | |
| E | I would like to work, but I am physically unable to. The state provides some benefits to help cover essentials, but I often struggle to pay the bills. | I am unable to visit family or friends, and I always feel lonely. I am very unhappy about my relationship status. | I would like to play sports but I am not physically able to do so. | |

*Figure 1 Life situations (vignettes) used for the survey.*

## 3.3. Gambles

To elicit individual utility functions, we employed a series of standard gamble tasks, structured around two core decision contexts:

1. **Personal risk decision (medical context):** Participants were asked to choose between a guaranteed life situation and a uncertain alternative with known probabilities of resulting in either a better or worse life outcome, or in some cases, death. The medical context was chosen for personal risk decisions, to facilitate participants to imagine a one-time, irreversible decision. Depending on the survey condition, options were either labelled (A-E) or defined by life satisfaction scores derived from previous vignette ratings.
2. **Societal inequality decision (policy-maker context):** Participants were asked to adopt the perspective of a policymaker choosing between two distributions of outcomes across a population. One policy offered equal life satisfaction outcomes for all, while the other produced unequal outcomes - benefiting some while disadvantaging others, potentially including death. To reinforce impartiality, participants were reminded that they themselves would be affected by the policy but would not know in advance whether they would benefit or lose out: “The policy will affect you as well, though you don’t yet know whether you will benefit or be negatively affected.”

Each participant completed the following three blocks of gambles, presented in randomised order within each block:

1. **Four personal risk gambles** between triples of adjacent life states on our ordinal scale, to estimate utilities for states A-E and death via chained standard gambles.
2. **Four societal inequality gambles**, again between triples of adjacent life states, to elicit utilities of states A-E and death via chained standard gambles.
3. **Four additional personal risk gambles** between randomly selected triples in which at least one state was non-adjacent, to allow testing the fit of the personal utility model through overidentification.

Personal risk gambles were presented first, as they were found in piloting to be more straightforward and hence familiarize participants with the task before presentation of the harder societal inequality gambles. The third block of non-adjacent gambles, deemed less central to the study, was positioned last to minimise the risk of survey fatigue during the core tasks.

Rather than ask participants for the exact probability at which they are indifferent to the gamble, we adopted a descending-probability procedure, in which gamble were first presented with a 1-in-2 chance of the worse outcome, followed by progressively lower probabilities: 1/5; 1/10; 1/100; 1/1,000; 1/10,000; 1/100,000; 1/1,000,000. If a participant refused a gamble, the next lower probability was shown. If they accepted a gamble (or refused the lowest possible odds) we proceeded to the next gamble. A 'can't choose' option was available throughout. If this option was selected, the next lower probability was offered to check whether the respondent was at the indifference threshold. If indecision persisted, the gamble was marked as undecidable, and the next gamble was presented. Participants could revise previous responses at any point.

To further reduce cognitive load, the user interface presented information that remained unchanged between questions in 'collapsed' form. Details of previous life satisfaction ratings, or of the same gamble with a higher set of odds, were not displayed unless requested. Visual animations highlighted which elements had changed between questions (e.g. odds only, or both options and odds).

Consistent with FDA recommendations for risk communication (Food and Drug Administration (FDA), 2016), the gambles were presented with pictograms to illustrate the probability of each outcome. In line with the principle of eliciting informed preferences (Frijters and Krekel, 2021) we also provided real-world comparators for each probability. For example: "For comparison, a UK adult aged 20–49 has a 1 in 100 chance of dying over a 10-year period."

Full examples of the gamble tasks are provided in the Supplementary Materials.

## 3.4. Analysis of standard gambles

For Objective 2, we derive both utilities and a loss aversion measure from gamble responses. Analysis of the standard gamble under EUM is based on Gafni (1994). If a participant *i* is indifferent to the choice of a fixed outcome, versus a gamble with probability $p_i$ of a worse outcome, then we can relate their personal utilities of the win, lose and baseline states ($U_{w,i}, U_{l,i}, U_{b,i}$) by Equation 1:

$$U_{b,i} = p_i U_{l,i} + (1 - p_i) U_{w,i} \tag{1}$$

For an ordered series of states, we define in ascending order for each participant $U_{Fi}, U_{Ei}, U_{Di}, U_{Ci}, U_{Bi}, U_{Ai}$. In the current study $U_F$ represents the utility of death. The utility scale is arbitrary in location and scale, so for the purpose of reporting results we fix two values $U_F = 0, U_A = 1$. However for estimation, we fix $U_F = 0, U_E = 1$ both to ensure a well-conditioned optimization, and to avoid the definition of $U_E, U_D, U_C, U_B$ as arbitrarily close to 1, where rounding errors can cause these utilities to appear equal in the presence of high levels of risk aversion. Utilities are rescaled between estimation and reporting. As we don't directly elicit a point of indifference from participants, we rely on observed choice data giving the highest known probability of failure at which each participant accepts each gamble (taken to be zero if they reject p=$10^{-6}$), and the lowest at which they reject it (taken to equal one if they accept the first gamble p=1/2). The point of indifference is therefore taken to be the midpoint of these values on a log scale.

Under CPT, a reference point must be specified, and losses and gains are then weighted differently with respect to this reference point. In a standard gamble framing, the certain outline $U_{b,i}$ is usually assumed to be the reference point (Oliver, 2003). The equivalent relation is then expressed as Equation 2:

$$0 = w_i^-(p_i)v_i^-\left(U_{l,i} - U_{b,i}\right) + w_i^+(1 - p_i)v_i^+\left(U_{w,i} - U_{b,i}\right) \quad (2)$$

Where $v_i^-, v_i^+$ are functions for each participant which allow for heterogenous valuation of losses and gains respectively, to capture loss aversion effects, while $w_i^-, w_i^+$ allow weighting of loss and gain probabilities respectively, to capture their over- or under-weighting of small probabilities. In the current study, however, we are considering only point estimates of utility rather than a continuous function, rendering loss and gain valuation functions unnecessary. Instead, we take $v_i^+(x) = v_i^-(x) = x$ and capture the relative valuation of losses compared to gains, $\lambda$, in a gamble between equal gains and losses - regardless of whether defined by CPT or EUM – by Equation 3:

$$\lambda = \frac{U_{b,i} - U_{l,i}}{U_{w,i} - U_{b,i}} \quad (3)$$

Noting that $0 < \lambda < \infty$ (with $\lambda = 1$ representing risk neutrality) we apply a log scale to create a symmetric measure centred on zero, followed by a standard logistic function to ensure finite bounds. In combination, these operations lead to Equation 4:

$$\lambda' = \frac{\lambda - 1}{\lambda + 1} \quad (4)$$

a measure in which $-1 < \lambda' < 1, \lambda = 0$ represents risk neutrality, and $\lambda = \pm x$ represents equal degrees of risk aversion/risk seeking on each side of the scale. When summarizing risk aversion tendencies, we take the mean value of $\lambda'$ over all relevant gambles.

The chained standard gamble gives only a point estimate for each item on a scale, with an exactly identified model lacking any quantification of error. We therefore also assess fit of the personal utility models by presenting further gambles, randomly selected from all gambles for states up to 3 steps apart on the ordinal scale, including gambles with asymmetric wins and losses. We fit the set of personal gambles for each participant using the binomial logit discrete choice model shown in Equation 5:

$$P\left(C_{i,gamble}\right) = \frac{\exp\left(\sigma_i U_{i,gamble}\right)}{\exp\left(\sigma_i U_{i,gamble}\right) + \exp\left(\sigma_i U_{i,baseline}\right)} \quad (5)$$

where $P(C_{i,gamble})$ is the probability of participant *i* choosing a given gamble with expected utility $U_{i,gamble}$ over a fixed outcome with utility $U_{i,baseline}$. $\sigma_i$ is a choice sensitivity parameter unique to the participant, estimated by maximum likelihood for each participant along with their personal utilities for states $U_{Di}, U_{Ci}, U_{Bi}, U_{Ai}$ (remembering we fix $U_F = 0, U_E = 1$ for estimation). Individual-level estimation is preferred over a random effects model, to fully visualise participant heterogeneity in utility curves; for this purpose we find the higher variance implicit in individual models to be acceptable.

For Objective 4, although the form of $w_i^-, w_i^+$ has high potential to affect results, we did not find any literature making empirical tests for values of these functions below p=0.01. Noting that the current study uses probabilities as low as p=0.000001, we can at best test sensitivity of our results to the extrapolated shape of $w_i^-, w_i^+$ defined in the literature. We base this on Gonzalez and Wu's (1999) test of probability weights at p=0.01 for the linear in log odds function given in Equation 6:

$$w_i^+(p) = w_i^-(p) = \frac{\delta_i p^{\gamma_i}}{\delta_i p^{\gamma_i} + (1-p)^{\gamma_i}} \tag{6}$$

where $\delta_i, \gamma_i$ are parameters unique to each participant; in analysis we test values for both Gonzalez and Wu's median participant, and their participant with greatest overestimation of small probabilities.

Analysis for all objectives is conducted in the Python/Jupyter ecosystem (Kluyver et al., 2016; Granger and Pérez, 2021) using SciPy (Virtanen et al., 2020), Pandas (Pandas development team, 2020; Wes McKinney, 2010), statsmodels (Seabold and Perktold, 2010) and Matplotlib (Hunter, 2007).

# 4. Results

## 4.1. Response validation

To assess comprehension and validate responses to the task, a supervised pilot was conducted with 20 participants via video call. This group demonstrated a good understanding of the survey instrument. Compared to the main, unsupervised sample, they took longer to complete the task (29 ± 11 vs. 23 ± 10 minutes; U = 3708, p = 0.02) and had slightly higher rates of failed attention checks (0.2 ± 0.4 vs. 0.07 ± 0.3; U = 3364, p = 0.03). However, there were no significant differences in risk tolerance, model fit (McFadden's $r^2$), the proportion of choices correctly predicted by the standard gamble model, or task completion rates.

Three responses were removed from the main unsupervised sample due to dropped server connections, yielding a final sample size of *n* = 297. Among these, 8% completed all societal gambles but not all personal ones, 4% completed all personal gambles but not all societal ones, and 5% were unable to fully complete either set. Feedback from these participants cited a lack of contextual information, difficulty making decisions on behalf of others, or discomfort with mortality-related choices. When analysis was restricted to non-lethal gambles, completion rates increased to 90%. No significant differences in completion rates were observed by political affiliation.

The distribution of life satisfaction ratings given to each situation is shown in Figure 2. Median life satisfaction ratings matched our own expectations for the imaginary scenarios, with A=10, B=8, C=6, D=4, E=2.

Some participants (5.1%) rated some scenarios out of order. One of these participants admitted an error in their response but did not correct it; one claimed it matched their experience of people they know; and the remainder provided no explanation when prompted. During piloting, some participants had provided an explanation for such ratings, expressing a need to experience challenge rather than a "perfect life". We discarded such responses as we consider them to represent a misunderstanding of what life satisfaction aims to measure.

No systematic differences were observed between the two survey conditions (life satisfaction before vs. after gambles) on any major outcome (completion time, risk tolerances, model fit).

Cronbach's alpha was 0.70 for the five-item political scale, indicating acceptable internal reliability for a short scale.

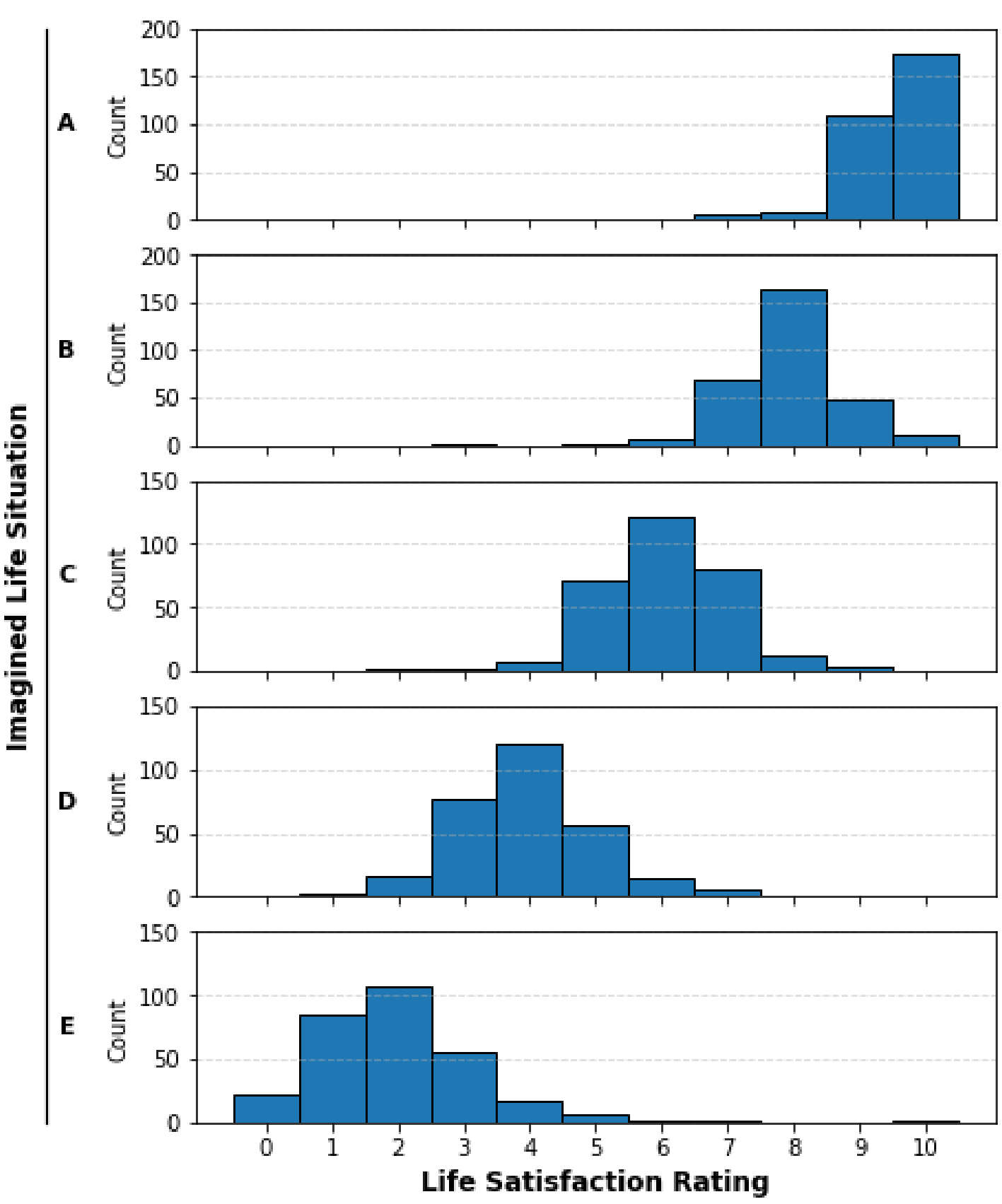

*Figure 2 Distribution of Life Satisfaction ratings for each imagined situation*

## 4.2. Estimation of personal and social utilities, and their relationship with political alignment

Estimated utilities of each situation are shown in Figure 3, (i) for personal decisions fitted both by chained standard gamble and by discrete choice, (ii) for societal/policy maker decisions fitted by chained standard gamble. Goodness of fit for each personal discrete choice model is characterized by McFadden's pseudo-$r^2$. 53% of participants have an individual choice model with pseudo-$r^2$>0.2, which is typically considered to represent good fit. The discrete choice model, fitted to 8 gambles, results in similar utilities to the chain of 4 gambles, though with greater smoothing and smaller variance between participants.

The societal utility curve is dominated by the high utility assigned to life itself, over any difference in life states. Some respondents in fact exhibit infinite risk aversion for societal gambles involving death. To capture their relative utilities for living states, it is therefore necessary to re-estimate utility curves excluding gambles with death.

We compute loss aversion ($\lambda$) for all gambles, with mean over gambles, and median over participants shown in Table 2. In all cases, the majority of participants demonstrate a clear aversion to both personal risk and societal inequality, with the latter being even more strongly avoided than the former. The aversion to societal inequality is most pronounced in gambles involving life-or-death outcomes (median $\lambda$=30.6) though still present in gambles between living states (median $\lambda$ for each gamble individually is 6.1). For personal choices, the median aversion to risk is lower for all states ($\lambda$=2.2). Importantly, neither of these tendencies shows any association with political alignment.

A plot of personal risk aversion versus social inequality aversion, for all gambles excluding death, is shown in Figure 4, along with the political alignment of each participant. The principal findings are clearly visible: 79% of participants showed aversion to personal risk, while 87% showed aversion to societal inequality.

Although a veil-of-uncertainty philosophy would predict equal aversion to societal inequality and personal risk, 73% showed an aversion to societal inequality which is greater than their own aversion to personal risk. The two variables are nonetheless clearly correlated, with people more willing to take risks themselves, also more likely to take risks for others ($r=0.62$, $p<0.01$).

These tendencies are independent of political alignment, in particular, the correlation of societal inequality tolerance and the political alignment scale is not significant ($r=-0.05$, $p=0.44$). The Mann-Whitney U-test, applied to voters for each political party, also shows no significant relationship with societal inequality tolerance, and only one significant result for personal risk aversion in a small sub-group (voting 'Other': n=4, mean 0.72±0.25 versus everyone else, n=268, mean=0.33±0.34; U-statistic 858, p=0.02).

| Gamble | $\lambda_{PERSONAL}$ (n) | $\lambda_{SOCIETAL}$ (n) | $\lambda_P>1$ | $\lambda_S>1$ | $\lambda_S \geq \lambda_P$ | $r(\lambda'_P,\lambda'_S)$ (p) | $r(\lambda'_P$,politics) (p) | $r(\lambda'_S$,politics) (p) |
|---|---|---|---|---|---|---|---|---|
| **E vs D/Death** | 2.2 [0.4, 30.6] (273) | 30.6 [2.2, 315.2] (269) | 66% | 84% | 86% | 0.46 (0.00*) | -0.03 (0.63) | -0.03 (0.61) |
| **D vs C/E** | 2.2 [0.4, 3.5] (278) | 6.1 [2.2, 30.6] (276) | 68% | 81% | 81% | 0.50 (0.00*) | 0.01 (0.84) | -0.04 (0.52) |
| **C vs B/D** | 2.2 [2.2, 6.1] (272) | 6.1 [2.2, 30.6] (276) | 74% | 83% | 81% | 0.47 (0.00*) | -0.07 (0.27) | -0.08 (0.20) |
| **B vs A/C** | 2.2 [2.2, 30.6] (280) | 6.1 [2.2, 315.2] (279) | 79% | 84% | 79% | 0.46 (0.00*) | -0.09 (0.12) | 0.02 (0.79) |
| **All gambles (phys health)** | 2.2 [1.0, 5.3] (271) | 4.7 [1.8, 30.6] (275) | 68% | 83% | 76% | 0.58 (0.00*) | -0.09 (0.13) | -0.04 (0.50) |
| **All gambles (no death)** | 2.2 [1.2, 4.0] (269) | 4.0 [1.8, 13.7] (272) | 79% | 87% | 73% | 0.62 (0.00*) | -0.07 (0.25) | -0.05 (0.44) |
| **All gambles** | 1.9 [1.1, 4.0] (263) | 4.4 [2.0, 15.9] (261) | 74% | 82% | 74% | 0.63 (0.00*) | -0.06 (0.29) | -0.04 (0.50) |

*Table 2 Median [Q1,Q3] (n) personal risk aversion ($\lambda p$) and societal inequality aversion ($\lambda s$) for all gambles between adjacent states. Percentage columns show proportion of participants exhibiting personal risk aversion, societal inequality aversion, and societal inequality aversion greater than personal risk aversion. Correlation columns use transformed values $\lambda'p$, $\lambda's$ and show Pearson's r (p-values) between personal risk aversion, societal inequality aversion and political alignment. Summary rows are derived from mean $\lambda'$ per participant, dropping participants who could not decide on all relevant gambles, then transformed back to $\lambda$ for which we report the median over participants. All gambles (phys health) refers to gambles between physically healthy states A, B, C and D only.*

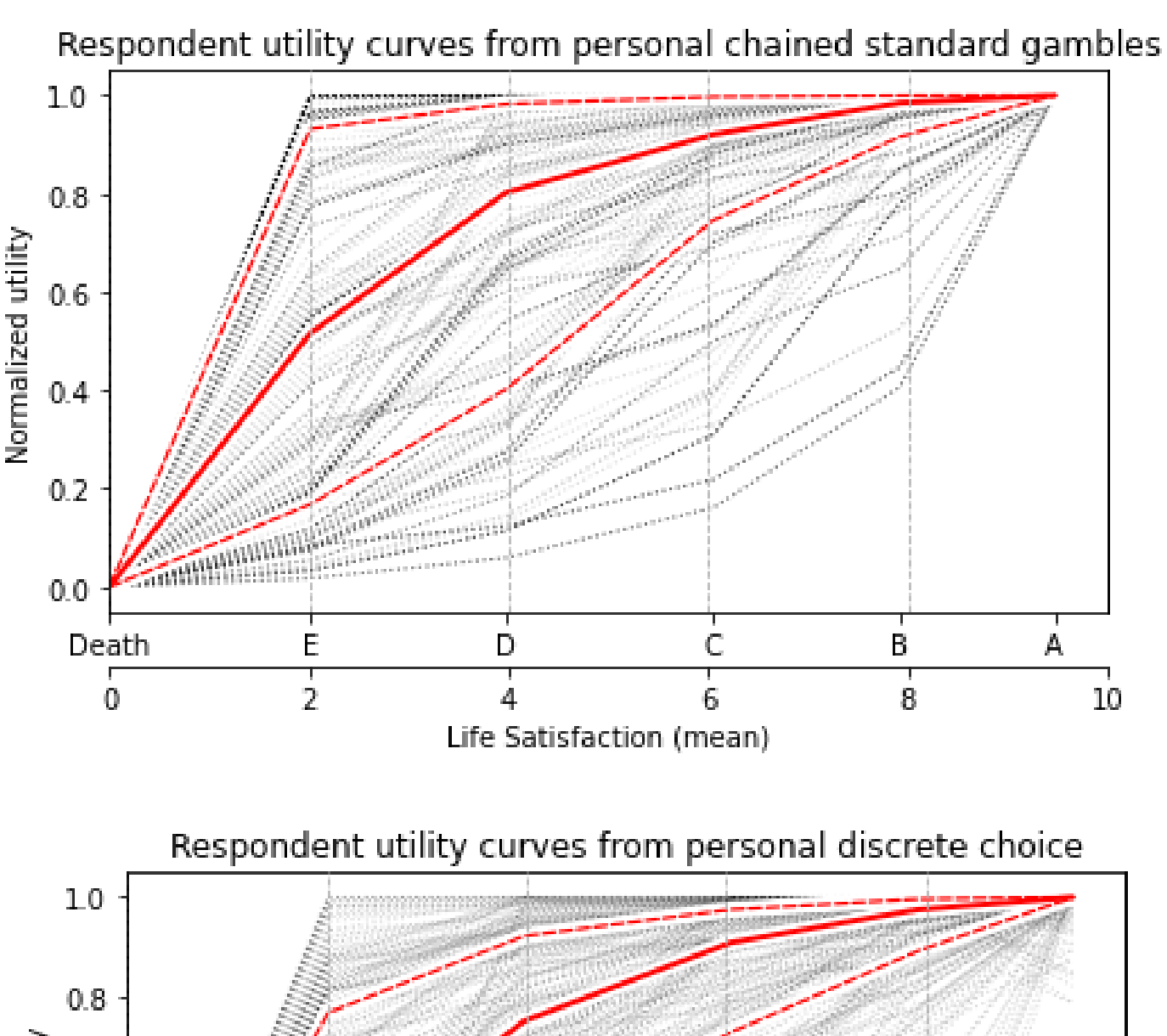

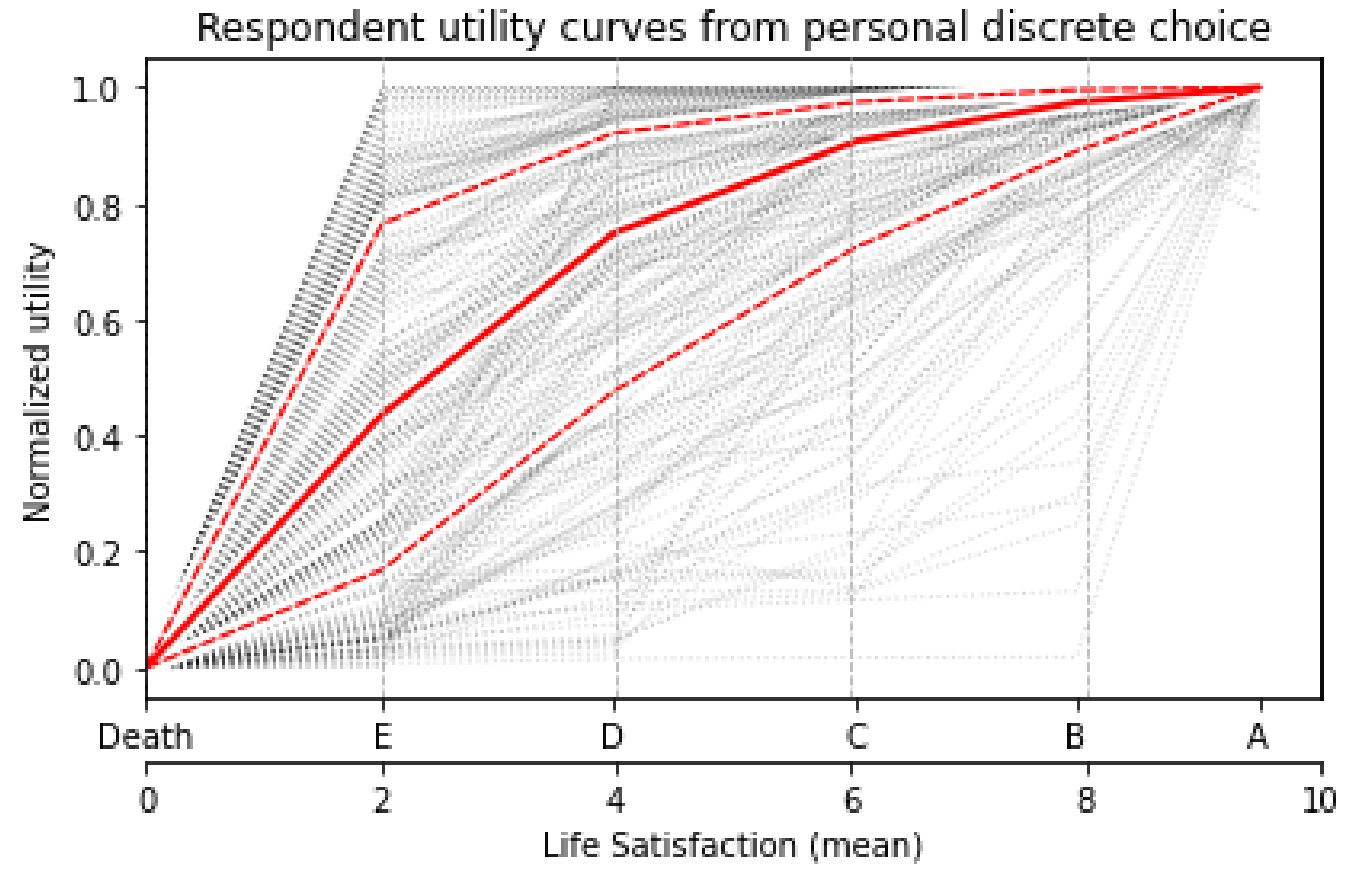

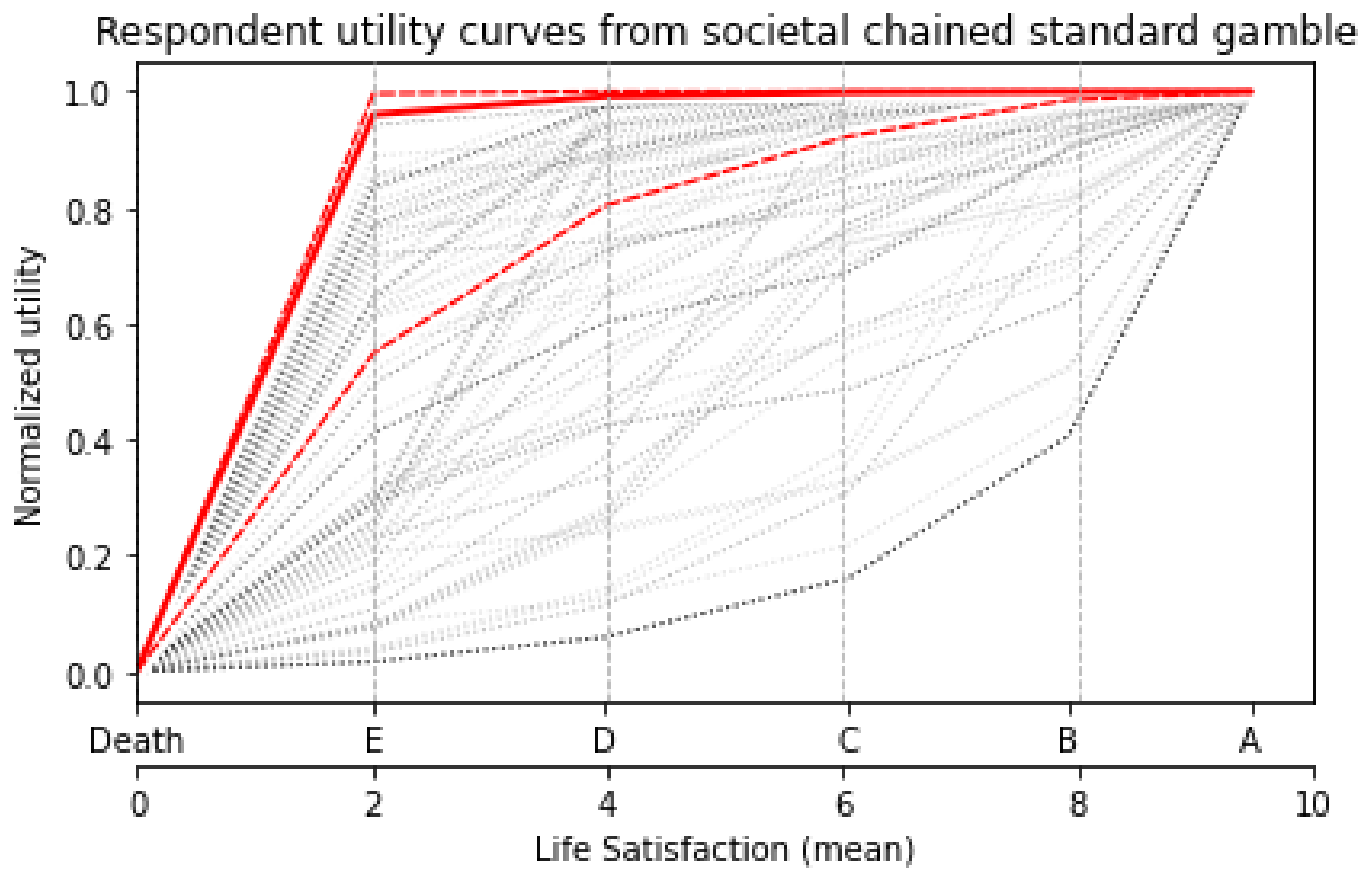

*Figure 3 Individual participant utility curves derived under EUM. Red lines show median utility and quartiles. X positions are plotted at the mean life satisfaction rating for each scenario (A-E) over all participants. From top to bottom: (1) Utilities derived from 4 personal risk choices using chained standard gamble. (2) The same utilities derived from discrete choice modelling of 8 personal risk choices. (3) Utilities derived from 4 societal inequality choices using chained standard gamble.*

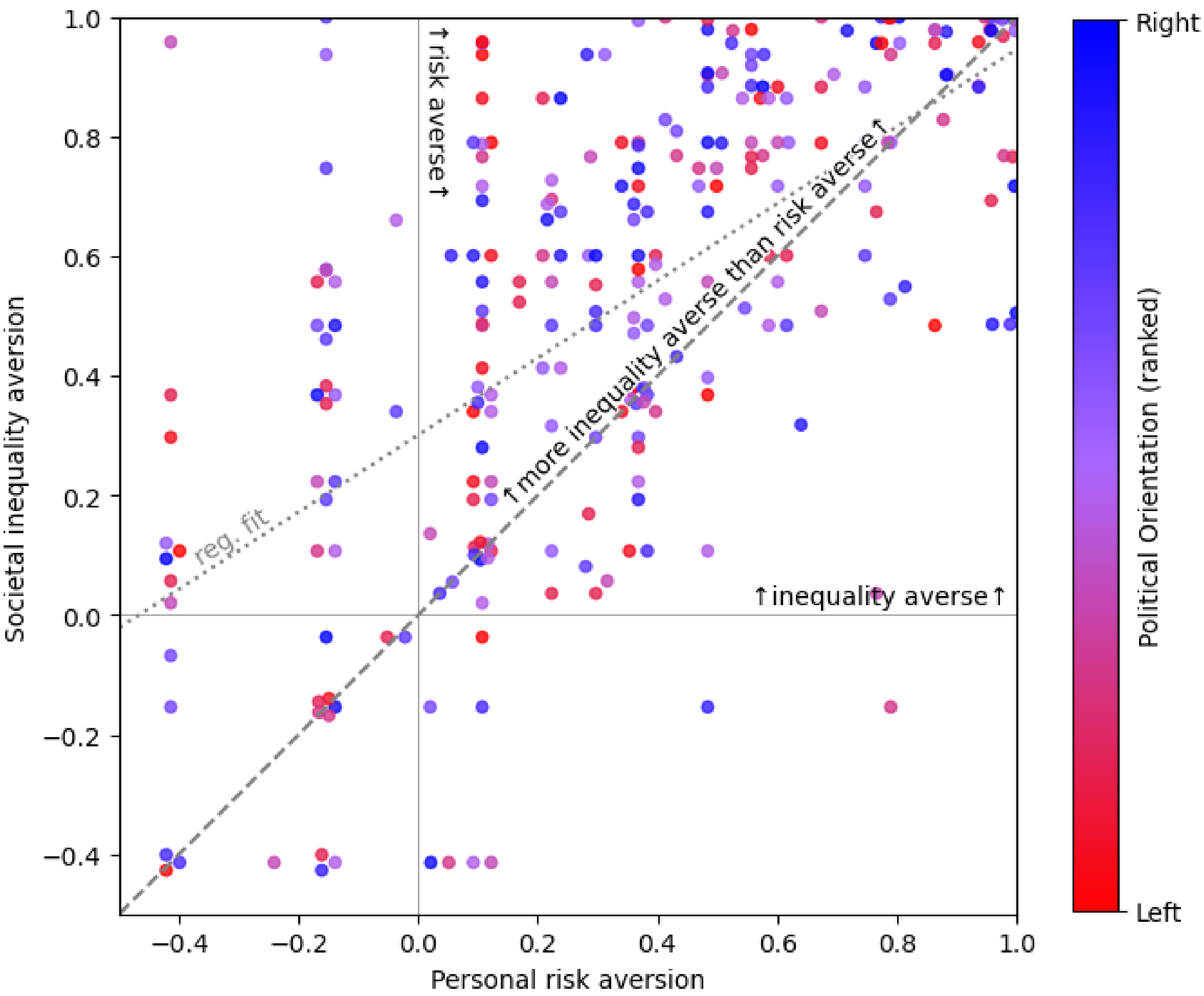

*Figure 4 Scatterplot of personal risk aversion versus societal inequality aversion, estimated from chained standard gamble choices (excluding those with risk of death) and coloured by political alignment. The diagonal line indicates a position consistent with the veil of uncertainty. Dotted line shows regression fit. 41 sets of coincident points (108/271 points in total) have been displaced by 0.015 units for display.*

## 4.3. Application of estimated utilities to national life satisfaction data

To illustrate the implications of these results, explore the use of our estimated utilities to summarize the distribution of life satisfaction in the UK (Office for National Statistics, 2023). The mean of the current life satisfaction distribution is 7.45, which as we argued in the introduction, is not representative of the variance due to the nonlinear utility of life satisfaction. We therefore compute Representative Life Satisfaction (RLS) levels – constant values which, if experienced by everyone, are equivalent to the distribution observed by the ONS based on (a) personal risk choices and (b) societal inequality choices. Equivalence is defined in two ways: (i) using median utility curves, giving the lowest RLS that 50% of participants would consider better than the distribution; (ii) using mean utility curves, giving the same total utility as the distribution, and thus taking account of strength of preference for each participant.

To allow interpersonal comparisons, individual utility curves are normalized by standard deviation using the current UK distribution of life satisfaction as the reference distribution. Under a democratic interpretation, this gives each individual utility curve equal "voting" power (Cotton-Barratt, MacAskill, and Ord, 2020). As the computation is sensitive to low utility states, for which ONS notes greater uncertainty in their estimates, the data has been binned to match ONS seasonally adjusted releases of life satisfaction data (Office for National Statistics, 2025) which defines ranges of 0-4, 5-6, 7-8, 9-10.

Results are shown in Table 3. RLS varies depending on the measure used, but for all four cases, is lower than the distribution mean.

| Utility curve | Representative RLS<br>(difference from mean LS of 7.45) |
|---|---|
| Personal risk, mean | 6.99 (-0.46) |
| Personal risk, median | 6.80 (-0.65) |
| Societal inequality, mean | 6.53 (-0.92) |
| Societal inequality, median | 5.69 (-1.76) |

*Table 3 Representative life satisfaction measures for the UK distribution of life satisfaction, 2023.*

## 4.4. Sensitivity to small probability weighting

We test sensitivity of the representative life satisfaction measures in Table 3, to application of probability weighting functions from CPT. Based on Gonzalez and Wu's (1999) linear in log odds function, with median parameter values across their participants ($\delta$ =0.77, $\gamma$=0.44), we would expect $w(10^{-6})=0.002$, overweighting these odds by a factor of 2,000. Applying this probability weighting in the above analysis, we find a recomputed version of Figure 4 to be qualitatively similar, albeit with all data points shrunk towards the origin. The reductions in RLS shown in parentheses in Table 3 reduce to the range [-0.30, -0.74]. Based on Gonzalez and Wu's single participant with the most extreme parameters ($\delta$=1.19, $\gamma$=0.27), we compute $w(10^{-6})=0.03$, overweighting these odds by a factor of 30,000. The recomputed Figure 4 is still qualitatively similar, albeit with further shrinkage towards the origin, and the reductions shown in parenthesis in Table 3 shrink to at most -0.36. We remind the reader in interpretation that this is a sensitivity test based on extrapolated figures from the cited study only, as probability weightings for $p<0.01$ have not been empirically measured in the literature.

# 5. Discussion

Our findings show a concave (risk-averse) valuation curve of well-being states, and therefore qualitatively align with both Peasgood (2018) and Mukuria (2023). However, we observe a higher degree of personal risk aversion than either study. This may be partly due to the overweighting of small probabilities; when this is accounted for in a sensitivity analysis, our estimates of personal risk aversion align more closely with those reported by Mukuria (excluding states worse than dead, which fall outside the scope of our study). In contrast, Peasgood reports milder risk aversion, potentially due to the lack of objective anchors for scale points, a feature incorporated in both Mukuria's study and our own. Notably, neither Peasgood nor Mukuria examined decision-making in interpersonal contexts. In such settings, our results indicate substantially greater societal risk aversion, suggesting individuals may adopt more cautious preferences when outcomes affect others rather than themselves.

Considering the representative life satisfaction measures computed, the largest RLS reduction can be interpreted as the average loss in life satisfaction that a typical participant might accept as a trade-off for achieving population-level equality in well-being - a striking figure of 1.8 points on the scale of 10. However, the preference difference between a 1.8-point and a 0.9-point life satisfaction reduction is marginal for the median participant. Under a democratic interpretation, where utility curves are normalized so that each individual curve has equal "voting" power (Cotton-Barratt, MacAskill, and Ord, 2020), most of the median "vote" is used to express strong preference for avoiding the lowest life satisfaction states. By definition this means expressing relative indifference between higher states, and a relatively high utility for the current state of the UK, where 94% of the population report a life satisfaction of 5 or higher. The mean utility approach captures not only the direction but also the intensity of individual preferences, resulting in less pronounced differences between mean life satisfaction and RLS. Even so, RLS based on mean utility still implies average willingness to trade away 0.9 points of personal life satisfaction in exchange for well-being equality, which is substantial.

Political orientation has long been known to be influenced by cultural and social factors, and for many voters is more a question of identity than policy preference (Converse, 2006). Nonetheless it is surprising to see no correlation whatsoever of social inequality aversion with politics. We suggest two possible reasons for this: firstly, that the posing of these questions in the life satisfaction domain – rather than the financial domain – may serve to decouple respondents' views from entrenched economic ideologies, partisan cues, and known financial reference points such as the participant's own income. Secondly, the fact that well-being outcomes are framed *without reference to its current social distribution,* removes any reference point allowing the participant to compare scenarios with their own political opinions on the status quo, and thus helps to uncover preferences on an absolute scale.

# 6. Strengths & Limitations

Although we have presented RLS measures which we propose are useful for summarizing distributions of well-being, we note that such measures quite sensitive to small changes in the distribution due to a double source of nonlinearity: (i) the transformation from life satisfaction to utility, and (ii) the transformation from utility back to life satisfaction, using a different parts of the same utility curve. When comparing multiple different policy outcomes, therefore, an average of normalized utility values is likely to be a better conditioned metric compared to RLS.

In a wider context, this evidence is highly relevant to the political questions of how governments should weigh alternative courses of action, and how future AI systems can be better aligned with human goals. While a full exploration of those questions is clearly beyond our scope, their relevance warrants a reflection on the limitations of interpreting our findings not only from an empirical, but also a normative perspective.

Taking a policy perspective, most participants exhibit risk aversion. Interpreting this tendency depends on a normative stance: whether such aversion reflects a cognitive bias in line with normative application of Cumulative Prospect Theory (CPT), or a legitimate expression of collective preference, consistent with Expected Utility Maximization (EUM). For both policymakers and AI systems supporting policymaking, whether this aversion is treated as a bias or a normatively valid preference will drastically affect their recommendations. Mischaracterizing it could lead to under- or over-weighting precaution in high-stakes scenarios, a critical issue for alignment.

In the context of political application, it should be noted that self-reported wellbeing measures are susceptible to bias from (i) the expectations individuals have for their lives, (ii) the incentive to misrepresent ones own well-being in order to influence policy (we note that within the current study, this latter limitation may have biased the reporting of 'zero' life satisfaction in the UK ONS data). Any attempt to use well-being measures in policymaking must therefore take account of these factors. Where unadjusted well-being measures are used, this relies on the reasonableness of people's life expectations being constant across all demographics. Aggregate approaches based on data collected outside of the political context, for interventions and life events having an equal effect on all demographics, are likely to be more reliable. This limitation also applies to AI training and evaluation datasets, where misreported well-being could introduce bias in aligned behaviour.

The current study focuses on outcomes and not process (ends and not means): in reality both are important. Future models could consider dynamic changes in well-being over time rather than a static endpoint, and also the effect of repeated gambles. This would also provide a framework for incorporating the effect of expectations on well-being.

The study is based on a UK sample and would benefit from internationalization. It would also benefit from presenting participants with more gambles on societal inequality, allowing for the fit of a discrete choice model to predict societal inequality tolerance as well as personal risk tolerance. Further nuance could be obtained by studying the effect of removing the veil of uncertainty, i.e., using a framing in which participants are told they will not be personally affected by a policy choice. However, given that we found most participants to be more sensitive to social inequality than to personal risk, we would not expect this change in framing to elicit substantially different results. Although our results remain qualitatively similar in the face of a sensitivity test, further research is recommended to experimentally test the degree to which participants over-weight the very small probabilities used in this study.

We did not present participants with life states considered worse than death. Although Peasgood et al. (2018)'s participants rate death as equivalent to approximately 2/10 on the life satisfaction scale, our own participants report high aversion to mortality risk even from a baseline of 2/10. More work is therefore needed to determine risk aversions between different low-satisfaction states, and between low-satisfaction states and death.

The reliance on stated preference is a final limitation. From an empirical, behaviourist perspective, future research could augment this with a revealed preference approach. However from a policy perspective, this same limitation is inherent in the democratic process itself. It may therefore be fruitful to compare existing consensus decisions made by the public, to the optimal-utility decisions predicted under this framework.

# 7. Conclusions

This study takes an empirical measurement of how people say they would balance risk, reward, equality and efficiency in well-being trade-offs concerning both themselves and others. From a political perspective, these findings are noteworthy. The traditional left-right political spectrum is often seen as capturing the degree to which individuals are willing to trade total wealth for more equal outcomes. The current study measures an analogous trade-off in the domain of well-being. We find that 87% of participants choose to make this trade-off to some extent, and, importantly, that the degree of trade-off each person prefers is unrelated to their political alignment.

In a time of sharp ideological divides over inequality, this suggests a unifying human intuition that transcends political ideology. Rather than attempting to reconcile polarised policy demands, decision-makers might instead build on shared intuitions about fairness in well-being as a basis for consensus.

With regard to metrics, the study shows that mean life satisfaction is not an adequate summary of well-being distribution. Representative Life Satisfaction (RLS) scores were found to be up to 1.8 scale points lower. We therefore propose the adoption of an RLS-based metric for national well-being reporting, and suggest adjustments to the WELLBY measure to account for the nonlinear utility effects observed in our data.

The question naturally arises as to which RLS measure to use. Although we have made progress in defining these measures based on public preference, we consider the final choice to require further research on consensus-building, as well as ethical discussion. Economists wishing to integrate our findings with a classical maximisation of utility will likely prefer a measure based on mean participant utilities, though an ethical question remains over whether they should be computed from personal or societal gambles, and a further consideration is the choice of reference distribution when normalizing utilities. Measures based on the median respondent bypass the issue of the reference distribution, but ignore strength of preferences.

Taking a broader view, the question of mean versus median RLS is a narrow attempt to answer how we should make decisions when individuals disagree. Given the wide range of risk tolerances exhibited by our participants, a fruitful avenue for future research will be to explore how groups can reach consensus decisions under such divergent preferences. We found that people report being more careful with the lives of others than with their own: some questions for future study are, why do we do this, and is it rational? Such care may simply reflect a precautionary principle learned in real conditions, where outcomes are less certain than in a stated preference experiment. Alternatively, it may reflect uncertainty in the preferences of others – in which case the differences between personal risk aversion, and societal inequality aversion, may disappear in the context of a group consensus-building exercise.

These findings also have implications for the long-term alignment of AI systems. The heightened interpersonal risk aversion we observe suggests that AI systems making decisions on behalf of others - especially across large populations - should adopt a more precautionary stance to better align with human intuitions about fairness and harm avoidance. The apparent independence of equity preferences from political identity may simplify the design of value-aligned systems, reducing the need to account for ideological divergence when modelling human values.

There are many ethical decisions which people think are wrong to make on the basis of utilitarian, numerical assessment of outcomes, whether or not adjusted for risk or inequality. Our findings do not change that. Human rights are considered to be inalienable, regardless of any utility calculation. The remit of a democratic government is limited by design, and overreach of powers should be avoided, regardless of any utility calculation. Further examples exist where utility calculations are widely considered to be inappropriate. For future AI systems, this highlights the importance of normative guardrails that cannot be violated, even in pursuit of aggregate gains. The authors recommend a strongly cautious stance, that any system needing these well-being utilities is not one which should be operating independently of human supervision – certainly not at the present time, and perhaps never.

Notwithstanding, there are many areas where decision-making by numbers is already widely accepted. Cost-benefit analysis, for instance, is generally seen as a legitimate framework, even if the details of specific calculations are subject to debate. Health economics - particularly in its valuation of human life - may attract more public controversy, yet it remains a well-established contributor to policymaking. In domains such as transport, computational simulations have long been used to support planning decisions rather than replace planners. In these, and similar areas, the present study offers both a methodological framework, and empirical grounding, to help ensure that the values embedded in numerical calculations better reflect the preferences of the wider public.

## Acknowledgements

We thank Dimitris Potoglou for feedback on the research design. This work is supported in part by the Clean Energy and Equitable Transportation Solutions (CLEETS) NSF-UKRI Global Centre award, under NSF award no. 2330565 and UKRI award no. EP/Y026233/1. For more information, please refer to https://www.cleets-global-center.org/

# Supplementary material: Sample survey questions

## 1. Personal risk gamble, life satisfaction basis

Now, consider the following scenario.

Imagine you are in a situation where you rate your life satisfaction as 8 out of 10.

Imagine that all your life, you have had a chronic health condition which you were born with, and which restricts your life somewhat. One day your doctor says you must choose between two treatments for this condition.

- Treatment A has a guaranteed outcome. You have thought about this outcome and you are certain that your life satisfaction after treatment would be 8 (out of 10).
- Treatment B is potentially better, but carries risk - the evidence shows that for some people it fails and makes the condition worse. You think that if the treatment succeeds, your life satisfaction would improve to 10, but if it fails, your life satisfaction would deteriorate to 6.

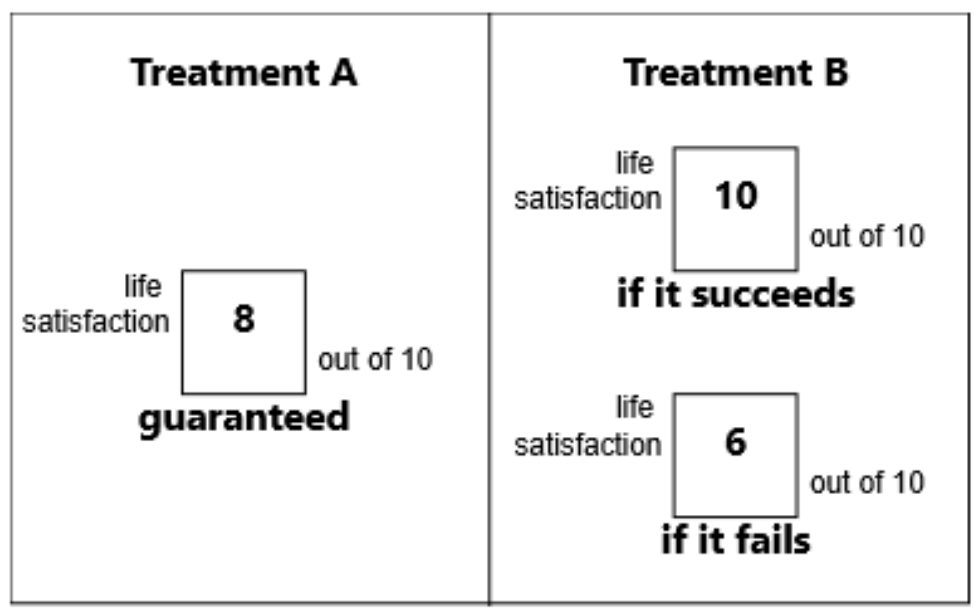

(*If you need to see how you rated different people's life satisfaction again, click* HERE)

---

The risks of Treatment B are well understood.

- It works perfectly for 1 out of 2 patients, but
- fails for 1 out of 2 patients.

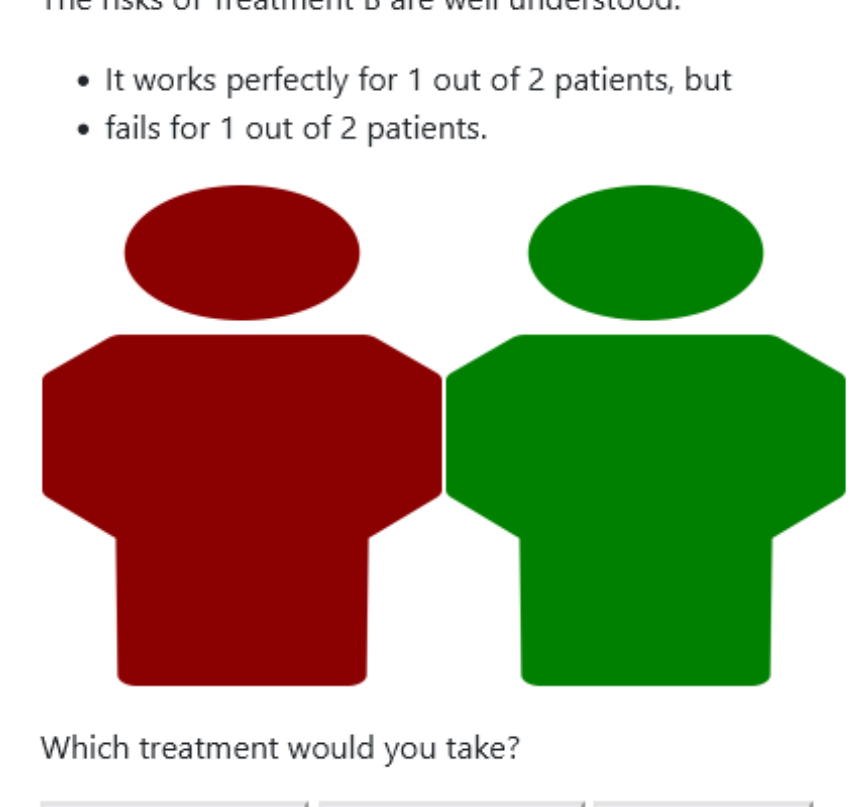

Which treatment would you take?

I would choose A | I would choose B | I can't choose

## 2. Policymaker scenario – life satisfaction basis

Now, consider the following **new** scenario.

**The context of this scenario is different - please read carefully!**

Imagine you are a **policymaker** who must make a choice affecting a large number of people. Currently, everyone in the affected group rates their life satisfaction as 2.

You have the following options:

- Policy A has a guaranteed outcome. Research shows that everyone's life satisfaction afterwards would be 2 (out of 10).
- The outcome of Policy B varies between people. Many people will benefit, and their life satisfaction afterwards would be 4 (out of 10). Others, however, would die as a result of the new policy. You cannot tell beforehand who would live or die.

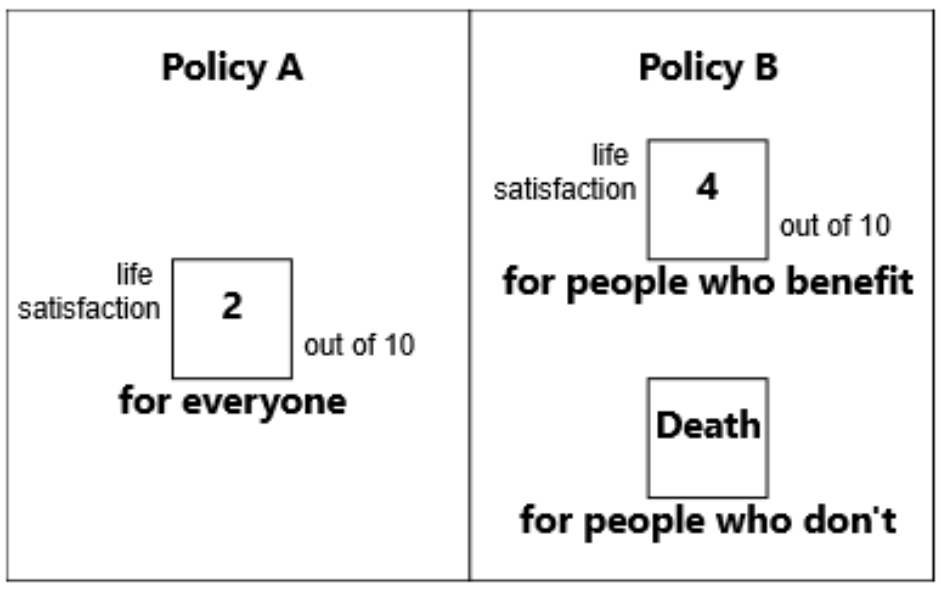

*You said you would choose Policy A.*

However, suppose the effects of Policy B were better, and it would

- benefit **99** out of **100** people, but
- the people it does not benefit (**1** out of **100** people) will die.

*The policy will affect you as well, though you don't know yet whether you will benefit or die.*

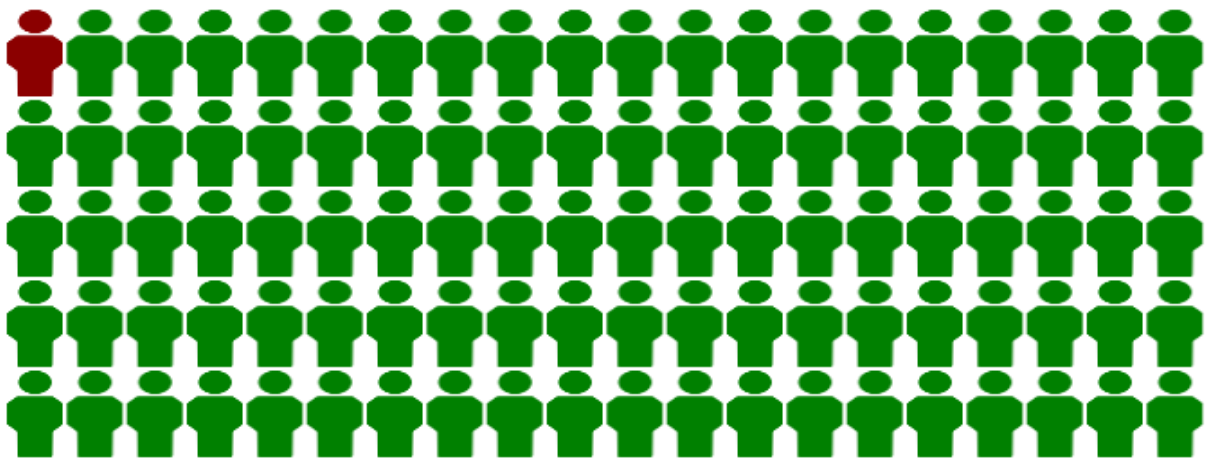

*For comparison, in everyday life, a UK adult (age 20-49) has a 1 in 100 risk of dying every 10 years.*

Which policy would you choose now?

| I would choose A | I would choose B | I can't choose | Go back (change previous answer) |
|---|---|---|---|

## 3. Personal risk scenario – vignette basis

Now, consider this scenario, based on the following life situations:

| Situation | Career | Relationships | Physical Fitness |
|---|---|---|---|
| D | I don't enjoy my job. The pay is low, and sometimes I struggle to pay the bills. | I only occasionally visit my family and friends, and I often feel lonely. I am unhappy about my relationship status. | I rarely exercise. I would like to exercise more but I struggle to find time and motivation. |
| E | I would like to work, but I am physically unable to. The state provides some benefits to help cover essentials, but I often struggle to pay the bills. | I am unable to visit family or friends, and I always feel lonely. I am very unhappy about my relationship status. | I would like to play sports but I am not physically able to do so. |

Imagine yourself living a life similar in overall quality to **Situation E**. *Your own life priorities may differ from Career, Relationships & Fitness, but please imagine yourself succeeding or struggling in areas that matter most to you, to a similar extent.*

Imagine that all your life, you have had a chronic health condition which you were born with, and which restricts your life somewhat. One day your doctor says you must choose between two treatments for this condition.

- Treatment 1 has a guaranteed outcome. Your life would remain comparable to **Situation E**.
- Treatment 2 has greater potential for improving your condition, although carries a risk of death. However if it succeeds, you think your life would improve to something like **Situation D**.

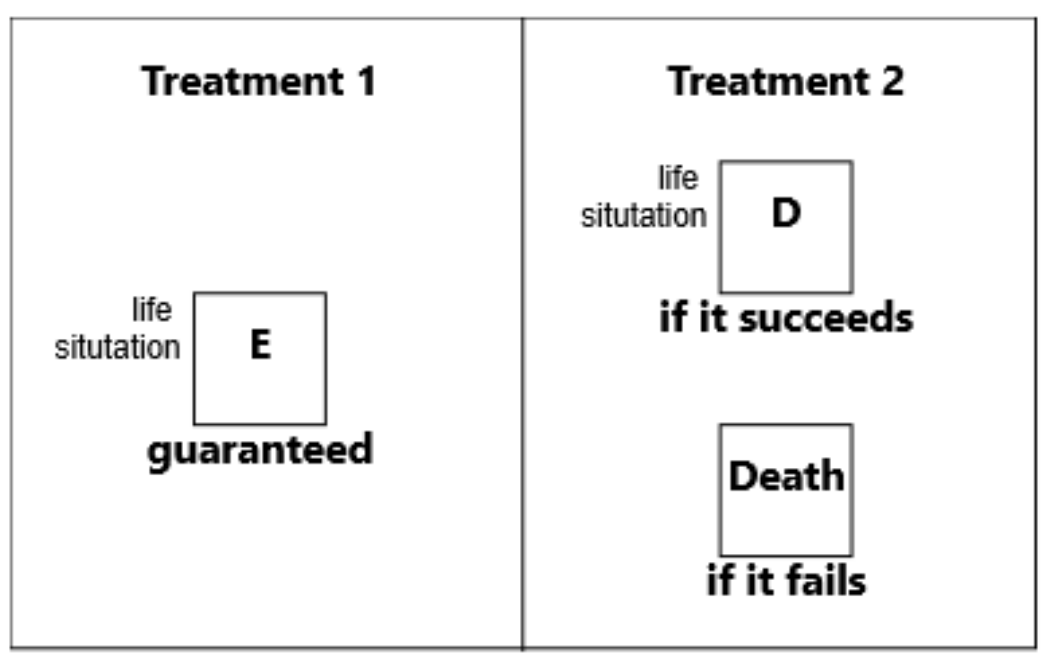

## 4. Political alignment

The political questions are as follows. “I am currently taking an online survey” is an attention check complying with the Prolific guidelines.

# Political opinions

We collect political opinion data to help us understand how people's political views affect their decision making.

Regarding politics, do you consider yourself more left leaning or right leaning? [Please select...]

What party are you most likely to vote for in the next election?
(If you are not eligible to vote, please select the option you would choose if you could)
[Please select...]

For each of the following statements, please select your level of agreement:

[Please select...] Government should redistribute income from the better off to those who are less well off

[Please select...] Big business benefits owners at the expense of workers

[Please select...] Ordinary working people do not get their fair share of the nation's wealth

[Please select...] I am currently taking an online survey

[Please select...] There is one law for the rich and one for the poor

[Please select...] Management will always try to get the better of employees if it gets the chance

[Next]